\documentclass[conference, 10pt]{IEEEtran}
\IEEEoverridecommandlockouts
\usepackage{algorithm}
\usepackage{algorithmic}
\usepackage{color}
\usepackage{bm}
\usepackage{graphicx}
\usepackage{amsmath}
\usepackage{amssymb}
\usepackage{multirow}
\usepackage{booktabs}
\usepackage{array}
\usepackage{amsthm}
\usepackage{lipsum}
\usepackage{enumerate}
\usepackage{stfloats}
\usepackage{cases}
\usepackage{halloweenmath}
\usepackage{cite}
\usepackage{subfigure}
\usepackage{pifont}
\usepackage{geometry}
\newcommand{\CN}{\mathcal{CN}}
\newcommand{\Expect}{\mathbb{E}}
\newcommand{\CC}{\mathbb{C}}

\title{An Adaptive Transmission Protocol Enabled by The State Switching Strategy of Beyond-Diagonal RIS}

\author{ Xiyao Wang\IEEEauthorblockN{$^{{1}}$ and Hongyu Li$^{2}$}

\IEEEauthorblockA{$^{1}$School of Information Science and Engineering, Harbin Institute of Technology (Weihai)  }

\IEEEauthorblockA{$^{2}$Internet of Things Thrust, The Hong Kong University of Science and Technology (Guangzhou) \\ Corresponding Author E-mail: \texttt{hongyuli@hkust-gz.edu.cn}  }
}

\def\BibTeX{{\rm B\kern-.05em{\sc i\kern-.025em b}\kern-.08em
    T\kern-.1667em\lower.7ex\hbox{E}\kern-.125emX}}
\begin{document}
\maketitle
\thispagestyle{empty}
\setlength{\columnsep}{0.201 in}

\author{\IEEEauthorblockN{1\textsuperscript{st} Given Name Surname}
\IEEEauthorblockA{\textit{dept. name of organization (of Aff.)} \\
\textit{name of organization (of Aff.)}\\
City, Country \\
email address or ORCID}
\and
}

\maketitle

\begin{abstract}
Thanks to inter-element interconnections and flexible element arrangements, the beyond diagonal reconfigurable intelligent surface (BD-RIS) breaks through the limitation of traditional RIS to achieve enhanced performance and enlarged coverage. However, existing BD-RIS research assumes that BD-RIS is always turned ON to assist transmission, while, in some scenarios where the transmitter-receiver direct link exists and remains strong, the performance gains of BD-RIS may not justify its manipulation complexity. In order to smartly use BD-RIS, a novel adaptive transmission scheme is proposed in this paper. The proposed scheme first defines two working states of the BD-RIS, namely ON and OFF. Based on these two states, we design a new beam training protocol that enables BD-RIS to intelligently switch its working state according to real-time channel conditions. Furthermore, we construct a decision threshold as the decision basis for state switching to ensure the efficiency and reliability of protocol execution. Simulation results show that, without always turning ON the BD-RIS and performing sophisticated wave manipulation, the proposed protocol can guarantee successful transmission regardless of whether there is a strong transmitter-receiver link or not. 
\end{abstract}


\section{Introduction}

Beyond diagonal reconfigurable intelligent surface (BD-RIS) has emerged recently as an upgrade of traditional RIS techniques, by introducing inter-element interconnections to work as a more intelligent relay in wireless environments with more flexibility for wave manipulation \cite{RIS2.0}.
Specifically, BD-RIS extends the functionality of traditional RIS through versatile architectures and modes, offering new degrees of freedom for the design of wireless systems \cite{RIS2.0}. Consequently, efficient and specialized channel acquisition and beamforming strategies are essential to adapt to the new constraints coming from the new hardware design and to fully exploit the potential of BD-RIS. Previous studies on conventional RIS have proposed various transmission protocols, two of which are commonly used, namely the channel estimation-beamforming-transmission protocol and the beam training-feedback-transmission protocol \cite{CE_RIS,CE1_RIS,CE_survey,FB,BT,HR}. 
The first protocol requires an explicit channel estimation phase, which is typically not easy for RIS-aided systems due to the passive property of RISs. 
The second protocol can be more practical since it replaces separate channel estimation and beamforming phases with beam training and feedback phases supported by predefined codebooks, thereby eliminating complicated channel acquisition and potentially improving the data transmission efficiency. 
As the key factor in the second protocol, the codebook construction should accurately reflect the mathematical constraints of RISs to facilitate the beam training. In this sense, previous studies effective for conventional RIS
are not directly applicable to BD-RIS due to its unique and more complex matrix constraints induced by inter-element connections.

Moreover, a critical and unresolved issue in BD-RIS deployment is the activation policy: \textit{Under which channel conditions should the surface be activated?} This question is especially pertinent in mixed (line-of-sight (LoS) and non-LoS (NLoS)) link environments. When a strong LoS channel component is present, the use of BD-RIS may introduce unnecessary beamforming complexity without substantial performance improvement. Conversely, in NLoS scenarios, it remains an open problem whether activating BD-RIS yields a net gain. Thus, developing an adaptive activation strategy is crucial for practical BD-RIS-aided systems.

Motivated by these considerations, we propose a novel transmission protocol for a multi-input single-output (MISO) system that smartly switches the working state of BD-RIS based on channel conditions. Specifically, we first define two BD-RIS operational states, namely ON and OFF, and design a new transmission protocol based on the state switching of BD-RIS. Then, within the proposed transmission protocol, we introduce a novel threshold-based method to detect the presence of a LoS channel from the transmitter to the receiver and propose a heuristic method to optimize the threshold.
Finally, we provide simulation results to validate the accuracy of the proposed threshold-based method and the effectiveness of the proposed protocol.

\textit{Notations:}
$\mathbb{C}$ denotes the set of complex numbers.
$(\cdot)^\mathsf{T}$ and $(\cdot)^\mathsf{H}$ denote the transpose and Hermitian, operations, respectively.
$\|\cdot\|_2$ denotes the $\ell$-2 norm of a vector and the Frobenius norm of a matrix.
$\otimes$ denotes the Kronecker product. $\mathbf{0}$ denotes an all-zero matrix with proper dimensions and $\mathbf{I}_M$ denotes an $M\times M$ identity matrix.

\section{System Model}\label{SM}
\begin{figure}[t!]
    \centering
    \includegraphics[width=0.48\textwidth]{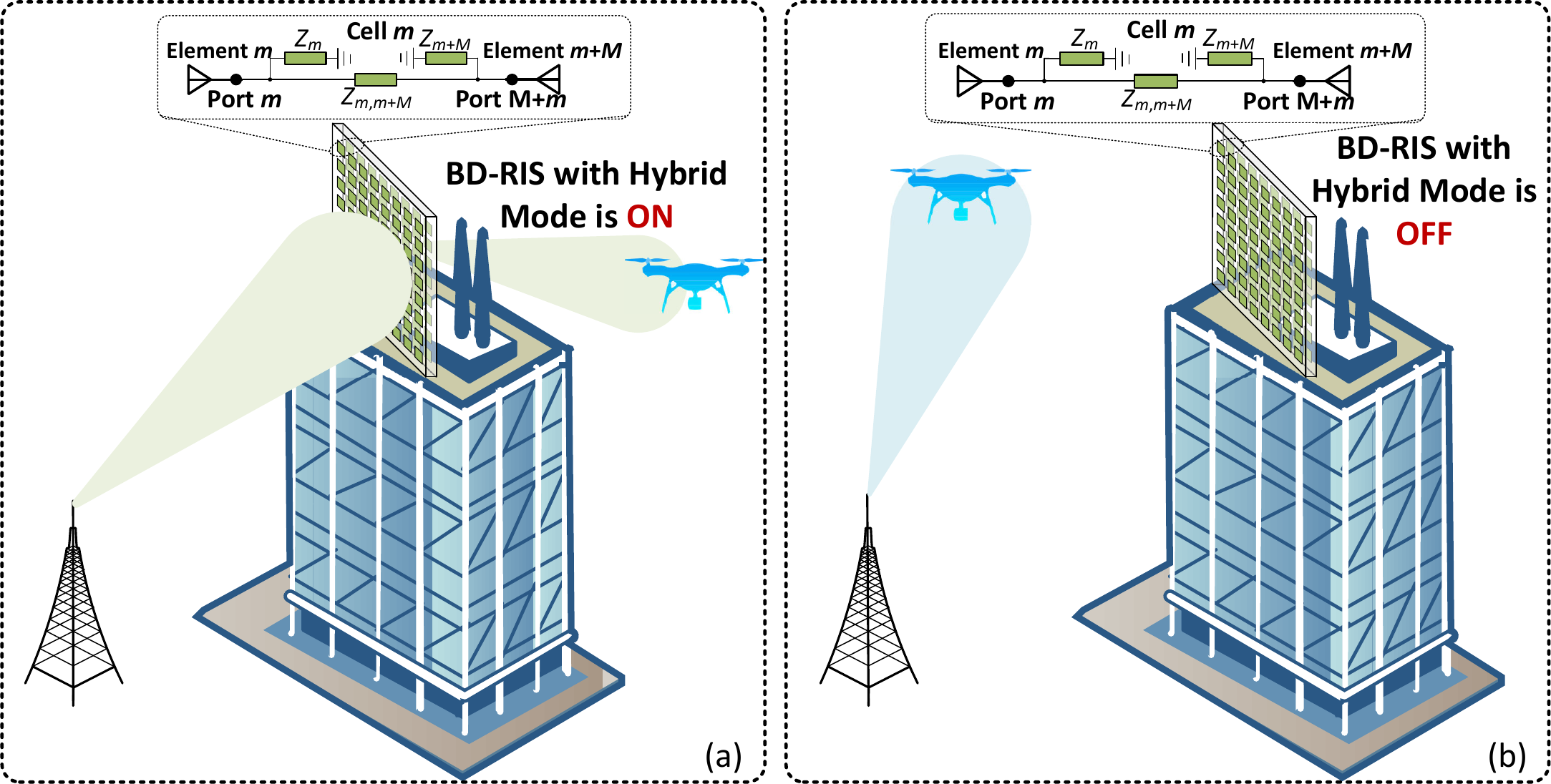}
    \caption{A MISO system where the BD-RIS works on either (a) the ON State or (b) the OFF State.}\label{model}
\end{figure}
\addtolength{\topmargin}{-0.04in}
Consider a downlink single-user MISO scenario, as illustrated in Fig.~\ref{model}. The base station (BS) is equipped with a uniform linear array (ULA) comprising $N_T$ antennas. The BD-RIS has $2M$ back to back placed elements with each side having $M$ elements \cite{RIS2.0}, and the user equipment (UE) has a single antenna. The elements from two sides of BD-RIS are interconnected with each other such that signals impinging on one side of the BD-RIS can be partially reflected from the same side and partially transmitted toward the other side. As such, the BD-RIS can support the hybrid transmitting and reflecting mode that includes purely transmitting and purely reflecting modes as two extreme cases. Due to the fixed deployment of BS and BD-RIS, and the available moving range of the UE, the LoS link from the BS to the UE does not always exist, as shown in Fig.~\ref{model}. Therefore, we propose the following BD-RIS state switching strategy:
\subsubsection {BD-RIS ON State} When the LoS path from the BS to the UE is obstructed, as shown in Fig.~\ref{model}(a), the BD-RIS is turned ON and works as a transmitting mode to establish a multi-hop channel to aid the data transmission. 
\subsubsection {BD-RIS OFF State} When the LoS path from the BS to the UE is unobstructed, as shown in Fig.~\ref{model}(b), the BD-RIS is turned OFF, with its scattering matrices being zero.

Below, we will first establish the transmission model based on two states of BD-RIS.

\subsection{Transmission Model}
\subsubsection{BD-RIS OFF State}
In this case, the received signal through the direct link is modeled as:
    $y = \sqrt{p} \mathbf{h}^\mathsf{H} \mathbf{x} + n$,
where $\mathbf{h} = \mathbf{h}_\text{LoS} + \mathbf{h}_\text{NLoS} \in \CC^{N_T}$  consisting of LoS and NLoS components, represents the channel from the BS to the UE, $n$ denotes additive white Gaussian noise following a complex Gaussian distribution $n \sim \CN(0,\sigma^2)$.
The transmit vector $\mathbf{x}$ is defined as $\mathbf{x} = \bm{w} s$, where $\bm{w} = [w_1, \cdots, w_{N_T}]^\mathsf{T}$ denotes the beamforming vector satisfying $\|\bm{w}\|_2^2 = 1$, $p$ is the transmit power, and $s$ represents a pilot symbol for beam training or a data symbol for data transmission, satisfying $\Expect[|s|^2] = 1$. 

\subsubsection{BD-RIS ON State}
In this case, the LoS path from the BS to the UE is blocked with $\mathbf{h}_\text{LoS} = \mathbf{0}$ and the BD-RIS is turned ON to establish a multi-hop link. A hybrid mode BD-RIS is subject to the matrix constraint \cite{RIS2.0}
$\mathbf{\Theta}^\mathsf{H}_\mathsf{r}\mathbf{\Theta}_\mathsf{r} + \mathbf{\Theta}^\mathsf{H}_\mathsf{t}\mathbf{\Theta}_\mathsf{t} = \mathbf{I}_M$,
where $\mathbf{\Theta}_\mathsf{r} \in \CC^{M \times M}$ denotes the reflection matrix and $\mathbf{\Theta}_\mathsf{t}\in \CC^{M \times M} $ denotes the transmission matrix of the BD-RIS. Based on the considered scenario, when the BD-RIS is turned ON, it always works on a transmitting mode. Therefore, we have $\mathbf{\Theta}_\mathsf{r} = \mathbf{0}$ and the received signal is given by
    $y = \sqrt{p}({\mathbf{h}}_2^\mathsf{H} {\bm{\Theta}_\mathsf{t}} {\mathbf{H}}_1 + \mathbf{h}_{\text{NLoS}}^\mathsf{H})\mathbf{x} + n$,
where ${\mathbf{H}}_1\in\CC^{M \times N_T}$ denotes the channel from the BS to BD-RIS and ${\mathbf{h}}_2 \in \CC^M$ denotes the channel from BD-RIS to the UE. 
The BD-RIS has a group-connected architecture where the elements are uniformly divided into $G$ groups and elements within the same group are connected. Therefore, we have a block-diagonal BD-RIS matrix satisfying 
\begin{equation}\label{eq:con_bdris}
    \bm{\Theta}_{\mathsf{t}}=\text{blkdiag}(\bm{\Theta}_{\mathsf{t},1},\ldots,\bm{\Theta}_{\mathsf{t},G}),~\bm{\Theta}_{\mathsf{t},g}^\mathsf{H}\bm{\Theta}_{t,g} = \mathbf{I}_{G_M},  \forall g,
\end{equation}
where $G_M = \frac{M}{G}$ denotes the group size \cite{RIS2.0}.

\textbf{Remark 1:} Note that simultaneous transmitting and reflecting (STAR) RIS \cite{STAR} is essentially a special case of the hybrid mode BD-RIS with $G_M = 1$. In this case, $\mathbf{\Theta}_\mathsf{r}$ and $\mathbf{\Theta}_\mathsf{t}$ become diagonal matrices and provide limited wave manipulation flexibility.  Simulation results will later show that BD-RIS with a group-connected architecture achieves better performance than STAR-RIS due to additional off-diagonal entries in $\mathbf{\Theta}_\mathsf{r/t}$. 


\subsection{Channel Model}
\subsubsection{BD-RIS OFF State}
The utilization of a horizontally configured ULA and narrow-beam transmission at the BS effectively suppresses multipath components. Therefore, the channel between the BS and the UE can be modeled by Rician fading with a high Rician factor, accurately characterizing its strong LoS-dominant propagation characteristics. Thus, the BS-to-UE channel is given by $\mathbf{h} =\mathbf{h}_\text{LoS} + \mathbf{h}_\text{NLoS} =\sqrt{\frac{K}{K+1} }\mathbf{h}_\text{L} + \sqrt{\frac{1}{K+1}}\mathbf{h}_\text{N}$,
where $K$ denotes the Rician factor, $\mathbf{h}_\text{L}=\sqrt{N_T}\alpha_1\bm{u}(N_T,\phi_d)$ denotes LoS channel component, with $\alpha_1$ capturing the path loss and $\bm{u}(N, \phi) = \frac{1}{\sqrt{N}} [1, e^{j\pi\beta\phi}, \ldots, e^{j(N-1)\pi\beta\phi}]^\mathsf{T}$ representing the normalized steering vector. 
Here, $\phi_d = \cos(\theta_d)\sin(\vartheta_d)$, where $\theta_d$ and $\vartheta_d$ are the azimuth and elevation angles of departure (AoD) for the direct path, respectively. The parameter $\beta = \frac{2d}{\lambda}$ is the wavenumber, where $d$ is inter-antenna spacing and $\lambda$ is wavelength. 
The vector $\mathbf{h}_{\text{N}}=[h_{\text{N},1},\cdots,h_{\text{N},N_T}]^\mathsf{T}$ represents the NLoS channel component with $h_{\text{N},i} \sim \CN(0,\alpha_1^2)$. 

In practical BD-RIS hardware, achieving a completely zero scattering matrix is challenging. Some approaches can be considered to approximate the OFF state. First, the RIS elements can adjust the equivalent impedance to a high reflection loss state. Second, impedance-matching structures can be integrated to reduce residual scattering.

\subsubsection{BD-RIS ON State}

For the obstructed LoS case, we assume that the BD-RIS is arranged as a uniform planar array (UPA) and model the BD-RIS channels in both horizontal and vertical dimensions. The BS-to-RIS channel ${\mathbf{H}}_{1}$ is modeled as 
${\mathbf{H}}_{1} = \sqrt{\frac{K}{K+1} }\mathbf{H}_{\text{L}1} +\sqrt{\frac{1}{K+1} }\mathbf{H}_{\text{N}1}$,
where $\mathbf{H}_{\text{L}1} = \sqrt{N_TM}\alpha_2 \bm{u}(\bar{M}, \phi_t) \otimes \bm{u}(\bar{M}, \varpi_t)\cdot\bm{u}^\mathsf{H}(N_T,\phi_t)$ denotes the LoS channel component. 
Here, $\phi_t = \cos(\theta_t)\sin(\vartheta_t)$, where $\theta_t$ and $\vartheta_t$ are the azimuth and elevation AoD of the BD-RIS path, respectively, $\bar{M} = \sqrt{M}$ is the number of elements in vertical and horizontal columns, $\alpha_2$ is BS-to-RIS path loss,  
$\varpi_t = \cos(\vartheta_t)$. 
$\mathbf{H}_{\text{N1}}\in \CC^{M \times N_T} $ is the NLoS channel component with $ \CN(0,\alpha_2^2)$.
Similarly, the RIS-to-UE channel is modeled as 
$\mathbf{h}_2 = \sqrt{\frac{K}{K+1} }\mathbf{h}_{\text{L2}}  +\sqrt{\frac{1}{K+1} }\mathbf{h}_{\text{N}2}$,
where $\mathbf{h}_{\text{L2}}=\sqrt{M}\alpha_3 \bm{u}(\bar{M}, \phi_r) \otimes \bm{u}(\bar{M}, \varpi_r)$ 
denotes the LoS channel component from BD-RIS to the UE, with
 $\phi_r = \cos(\theta_r)\sin(\vartheta_r)$ and $\varpi_r = \cos(\vartheta_r)$. 
Here, $\theta_r$ and $\vartheta_r$ are the azimuth and elevation angle of arrival (AoA) for the transmission path, $\alpha_3$ is RIS-to-UE path loss.
$\mathbf{h}_\text{N2}=[h_{\text{N}2,1},\cdots,h_{\text{N}2,M}]^\mathsf{T}$ is the NLoS channel component from RIS to UE with $h_{\text{N}2,i} \sim \CN(0,\alpha_3^2)$. 
\addtolength{\topmargin}{-0.03in}

\subsection{Problem Description and Simplification}
Through the above modeling, we aim to achieve efficient data transmission by properly selecting the BD-RIS state of the considered system.
Note that when the BD-RIS is ON, the BD-RIS matrix and BS beamforming vector are coupled. To decouple the joint optimization, we propose a simplified design for $\bm{w}$. Given that the channel $\mathbf{H}_1$ is often distributed with a strong LoS component and the locations of both BS and BD-RIS are static, we can avoid the need to know the exact information of $\mathbf{H}_1$. Instead, according to the principle of the matched filter \cite{match_filter}, we efficiently set the beamforming vector to align with the geometric LoS direction between the BS and the BD-RIS, i.e., $\bm{w}_0 = \bm{u}_1(N_T,\phi_t)$. 
In this case, we have $\mathbf{H}_1\bm{w}_0 = 
{\mathbf{h}}_{1e} = \sqrt{\frac{K}{K+1}}\mathbf{h}_{\text{L}1e}+\sqrt{\frac{1}{K+1} }\mathbf{h}_{\text{N}1e}$,
where $\mathbf{h}_{\text{L}1e} = \sqrt{N_TM}\alpha_2 \bm{u}_1(\bar{M}, \phi_t) \otimes \bm{u}_2(\bar{M}, \varpi_t)$ and $\mathbf{h}_{\text{N}1e}= \mathbf{H}_{\text{N}1}\bm{u}_1(N_T,\phi_t)$.

Then, we evaluate the system performance by the effective throughput, which takes into the tradeoff between the achievable rate and the training overhead. Let $T_c$ denote the coherence block length measures in the symbols. We use $\tau_\text{OFF}$ and $\tau_\text{ON}$ to denote the transmission overhead of the BD-RIS OFF and ON state, respectively. Their specific values depend on the beam training codebook size which will be introduced in Section~\ref{PD}.  
When the BD-RIS is OFF, the system transmits data through the direct link. For a selected BS beamforming vector $\bm{w}$, the OFF state signal-noise-ratio (SNR) is $\gamma_{\text{OFF}} = \frac{p|\mathbf{h}^{\mathsf{H}}\bm{w}|^2}{\sigma^2}$. When the BD-RIS is ON, the system transmits through the blocked direct link and the reflect link. By aligning $\bm{w}$ to $\mathbf{H}_1$ and selecting a BD-RIS matrix $\bm{\Theta}_{\mathsf{t}}$, the ON state SNR is     
$\gamma_{\text{ON}}
    =
    \frac{
    p\left|
    \mathbf{h}_2^{\mathsf{H}}\bm{\Theta}_{\mathsf{t}}\mathbf{h}_{1e}
    +
    \mathbf{h}_{\mathrm{NLoS}}^{\mathsf{H}}\bm{w}_0
    \right|^2
    }{\sigma^2}$.
The effective throughput maximization problem is 
\begin{equation}
\label{eq:simplified}
\left\{
\begin{aligned}
R_{\mathrm{OFF}}^{\star}
&= \max_{\bm w}\; \Big(1-\frac{\tau_{\mathrm{OFF}}}{T_c}\Big)\log_2\!\left(1+\gamma_{\mathrm{OFF}}\right)
\; \mathrm{s.t.} \; \|\bm w\|_2^2=1,\\[1mm]
R_{\mathrm{ON}}^{\star}
&= \max_{\Theta_t}\;
\Big(1-\frac{\tau_{\mathrm{ON}}}{T_c}\Big)\log_2\!\left(1+\gamma_{\mathrm{ON}}\right)
\; \mathrm{s.t.} \; \eqref{eq:con_bdris}.
\end{aligned}
\right.
\end{equation}

Directly solving problem  \eqref{eq:simplified} requires explicit channel state information (CSI) for channels $\mathbf{h}$, $\mathbf{h}_{1e}$, and $\mathbf{h}_2$. 
Fast channel acquisition for BD-RIS aided scenarios is unfortunately not easy due to the passive property and unique architecture of BD-RIS \cite{BD_RIS_CE}.  Therefore, the system requires the design of a low-overhead and adaptive beam training protocol. This protocol must achieve fast beam alignment between the BS and the UE (via either the direct link or the BD-RIS link), intelligently switch between the two operation states (BD-RIS OFF and ON), and exhibit high reliability in detecting LoS link blockage and executing state decisions. Furthermore, the protocol should operate with limited feedback to avoid the need for full channel estimation, which is computationally expensive and time-consuming.

\section{Protocol Design}\label{PD}

In this section, we propose a novel transmission protocol based on the state switching strategy of BD-RIS. 

\subsection{Protocol}
\begin{figure}[t!]
    \centering
    \includegraphics[width=0.48\textwidth]{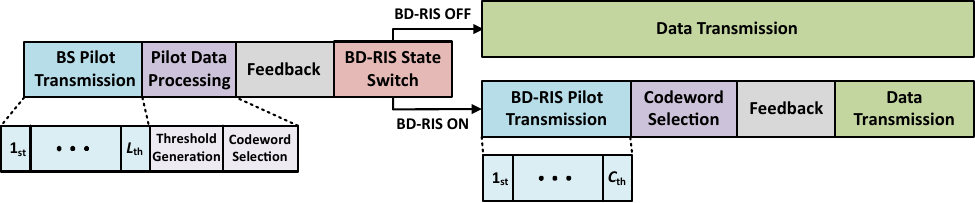}
    \caption{Diagram of the proposed transmission protocol.}\label{protocol}\vspace{-0.5 cm}
\end{figure}

Through the above modeling and discussion in Section \ref{SM}, we propose a transmission protocol as shown in Fig.~\ref{protocol} and summarize as follows.

\textbf{Phase 1:} The BS first attempts direct link communication using a codebook to transmit multiple training pilots. The UE measures the power of each codeword. If any power value exceeds a certain threshold, the UE feeds the index of the optimal codeword to BS and jumps to \textbf{ Phase 3}. If all power values are below the threshold, the UE feeds back a zero and proceeds to the next phase.

\textbf{Phase 2:} The BS performs the matched filter to align with the BD-RIS and obtain the effective channel $\mathbf{h}_{1e}$, and then transmits multiple training pilots assisted by BD-RIS, during which the BD-RIS regularly changes its configuration according to a codebook. The UE feeds the index of the optimal codeword back to the BS and BD-RIS.

\textbf{Phase 3:} The BS (using the optimal codeword from \textbf{Phase 1} if the  LoS BS-UE path exists and using the matched filter if not) and BD-RIS (if \textbf{Phase 2} is used) employ the optimal codewords that represent the optimal beams for beam alignment and data transmission.

\addtolength{\topmargin}{-0.03in}
\subsection{Beam Training and Detection of the BS}

In the first phase of the protocol, the system detects and evaluates the state of the direct link through a beam training process. 
The BS employs a predefined beam codebook $\mathcal{W} = \{\bm{w}(1), \cdots, \bm{w}(L)\}$, where $L$ is the number of codewords. It sequentially selects each codeword $\bm{w}(l)$ (for $l \in \{1, \cdots, L\}$) to transmit a training pilot. 
Here we simply use the well-known discrete Fourier transform (DFT) matrix to construct an orthogonal codebook $\mathcal{W}$.
The codebook not only fulfills the orthogonality requirement essential for threshold derivation analysis in Section \ref{TD}, but covers the whole spatial domain to facilitate effective beam scanning. 
The corresponding received signal for each codeword is given by 
\begin{align}\label{transbm}
    y (l)= \sqrt{p} \mathbf{h}^\mathsf{H} \bm{w}(l) s(l) + n(l), \forall l,
\end{align}
where we simply set $s(1) = \ldots = s(L) = 1$ and assume $n(l)\sim\mathcal{CN}(0,\sigma^2)$. 
The UE collects all $L$ received signals and generates the statistical information to get the decision threshold $T$. 
The UE also compares the power of the received signal to identify the optimal index $l^\star$ that leads to the maximum received power. 
By comparing $|y(l^\star)|^2$ with the threshold $T$ that will be derived in Section \ref{TD}, the UE makes a decision, that is if the power of the received signal exceeds $T$, it feeds back the index $l^\star$ to the BS; otherwise, it feeds back a zero.

\subsection{BD-RIS Assisted Beam Training}
In the obstructed LoS scenario (BD-RIS is turned ON and works on the transmission mode), we adopt and adapt the codebook structure proposed in \cite{BD_RIS_CE} to fit the specific system model considered in this model. This allows us to rewrite the BD-RIS assisted channel as 
\vspace{-0.08in}\begin{equation}\label{eq:channel}\vspace{-0.05in}
    \mathbf{h}_2^\mathsf{H}\mathbf{\Theta}_\mathsf{t}\mathbf{h}_{1e} = \sum_{g=1}^G(\mathbf{h}_{1e,g}^\mathsf{T}\otimes\mathbf{h}_{2,g}^\mathsf{H})\bm{\varphi}_g = \bm{f}^\mathsf{T}\bm{\varphi},
\end{equation}
where we define $\mathbf{h}_{2,g} = [\mathbf{h}_2]_{(g-1)G_M+1:gG_M}$, $\mathbf{h}_{1e,g} = [\mathbf{h}_{1e}]_{(g-1)G_M+1:gG_M}$, and $\bm{\varphi}_g = \text{vec}({\bm{\Theta}_{\mathsf{t},g}})$ with $\text{vec}(\cdot)$ being the vectorization operation. In addition, we introduce $\bm{f} = [(\mathbf{h}_{1e,1}^\mathsf{T}\otimes\mathbf{h}_{2,1}^\mathsf{H}),\ldots,(\mathbf{h}_{1e,G}^\mathsf{T}\otimes\mathbf{h}_{2,G}^\mathsf{H})]^\mathsf{T}$ and $\bm{\varphi} = [\bm{\varphi}_1^\mathsf{T},\ldots,\bm{\varphi}_G^\mathsf{T}]^\mathsf{T}$. From equation (\ref{eq:channel}) we observe that the configuration $\bm{\varphi}$ of BD-RIS tightly depends on the cascaded channel $\bm{f}$, which cannot be easily and efficiently acquired due to the passiveness of BD-RIS \cite{CE_survey}.  
Therefore, we perform a fast training procedure without exact CSI, as will be explained below. 

The BD-RIS utilizes a predefined configuration codebook $\mathcal{\varPhi} = \{\bm{\varphi}(1), \cdots, \bm{\varphi}(C)\}$, where $C$ is the number of codewords, and sequentially switches through each configuration vector $\bm{\varphi}(c)$, where $c \in \{1, \cdots, C\}$. 
Specifically, the codebook $\mathcal{\varPhi}$ is constructed following Lemma 1 and Theorem 1 in \cite{BD_RIS_CE} to guarantee the orthogonality between codewords.
We have the received signal corresponding to each codeword as
\begin{equation}\label{transrisbm}
    y(c) = \sqrt{p} [\bm{f}^\mathsf{T}\bm{\varphi}(c) + \mathbf{h}_\text{NLoS}^\mathsf{H}\bm{w}]s(c) + n(c), \forall c,
\end{equation} 
where we again set $s(1) = \ldots = s(C) = 1$ and assume $n(c)\sim\mathcal{CN}(0,\sigma^2)$. 
During the training process, the BS successively transmits pilot signals while the BD-RIS sequentially switches its configuration according to the codebook $\mathcal{\varPhi}$. The UE then compares the power of $C$ received signals, identifies the optimal codeword index $c^\star$ that leads to the maximum received power, and feeds it back to BD-RIS and BS.

With the above definitions of the BS codebook size $L$ and the BD-RIS codebook size $C$, the overhead introduced in Section~\ref{SM} can be specified\footnote{Since the overhead for codeword selection and feedback is negligible compared to that for pilot transmission, in this work we only capture the impact of pilot transmission overhead.}. If the BD-RIS is OFF after Phase 1, only the BS beam training is performed, thus $\tau_{\text{OFF}} = C$. If the BD-RIS is ON after Phase 1, the BS beam training and the BD-RIS beam training are performed, thus $\tau_{\text{OFF}} = L+C$. 


\section{Threshold Derivation and Optimization}\label{TD}
Threshold design is a fundamental aspect of the proposed protocol in Section \ref{PD}, directly impacting the reliability of link state detection. In codebook-driven transmission systems, accurate detection under varying channel conditions requires a robust threshold derivation method. This section presents a heuristic approach based on statistical signal processing to optimize the decision threshold and further enhance the system performance.

According to the Neyman-Pearson principle \cite{lehmann1993fisher} and the inclusion-exclusion principle, the accuracy probability $P_{\text{a}}$ maximization problem formulates as 
\begin{equation}\label{np_1}
    \begin{aligned}
        \max_{T}\quad &P_\text{a} = \mathbb{P}(\mathcal{H}_0)P_\text{d} + \mathbb{P}(\mathcal{H}_1)(1-P_\text{fa})\\
        \text{s.t.} \quad& P_{\text{fa}} = \mathbb{P}(|y(l^\star)|^2>{T} \mid \mathcal{H}_1),\\
        & P_{\text{d}}=\mathbb{P} (|y(l^\star)|^2>T \mid \mathcal{H}_0),
    \end{aligned}
\end{equation}
where $\mathcal{H}_0$ and $\mathcal{H}_1$ denote the cases in which the LoS channel component is present and not, respectively, $P_\text{fa}$ denotes the false alarm probability, and $T$ denotes the decision threshold. Problem (\ref{np_1}) is difficult to solve since both the objective function and constraints do not have explicit expressions with respect to the threshold $T$. To simplify optimization, we first derive the closed-form expression between $P_\text{fa}$ and $T$ below. 
\addtolength{\topmargin}{0.03in}
Under $\mathcal{H}_1$, the received signal contains only the NLoS component plus noise and is expressed as
    $y(l) = \sqrt{p}\mathbf{h}_{\text{NLoS}}^\mathsf{H}\bm{w}(l)s(l)  + n(l)$.
With $\mathbf{h}_{\text{NLoS}}\sim\mathcal{CN}(\mathbf{0},\frac{\alpha_{1}^2}{K+1}\mathbf{I}_{N_T})$ and $n(l)\sim\mathcal{CN}(0,\sigma^2)$, the received power $|y_{\text{NLoS}}|^2$ has a probability density function (PDF) following an exponential distribution as
$f(|y(l)|^2) = \frac{1}{\epsilon} \exp\left(-\frac{|y(l)|^2}{\epsilon}\right)$,
where $\epsilon = p \frac{\alpha_{1}^2}{1+K}  + \sigma^2$ denotes the variance of $y(l)$.
Then, the false alarm probability $P_\text{fa}$ is expressed as
\begin{equation}\label{Pfa}
\begin{aligned}
    P_\mathrm{fa} &= \mathbb{P}(|y(l)|^2 > T) \\
    &= \int_{T}^{+\infty} f(|y(l)|^2) \, d(|y(l)|^2).
\end{aligned}
\end{equation}
By solving \eqref{Pfa}, we can obtain the closed-form expression of $T$ in this case
as
\begin{equation}
T = -\ln(P_{\text{fa}})\epsilon = -\ln(P_{\text{fa}})\left( \frac{p\alpha_{1}^2}{1+K}  + \sigma^2\right).\label{eq:T_closedform1}
\end{equation}

Under $\mathcal{H}_0$, the received signal shares the same variance $\epsilon$ as that under $\mathcal{H}_1$ (a detailed derivation is summarized in the Appendix). This implies that we can directly collect the received signals in the environment, regardless of whether the LoS exists or not, calculate received powers, and estimate $\epsilon$ accordingly. In practice, the parameter $\epsilon$ is not fixed and can vary with channel states, while the threshold depends heavily on the value of $\epsilon$. Therefore, we propose a dynamic threshold by estimating $\epsilon$ and replacing it into \eqref{eq:T_closedform1} given a fixed $P_\mathbf{fa}$, which will be elaborated below.

Based on the above derivation, the threshold $T$ is uniquely determined by the false alarm probability $P_\text{fa}$ such that $P_\text{a}$ in (\ref{np_1}) is essentially a function of $P_\text{fa}$.
Therefore, we perform an offline calibration to select a suitable false alarm probability $P_{\text{fa}}^\star$ that maximizes $P_\text{fa}$.
Specifically, the BS constructs a calibration dataset that has $I$ representative samples (e.g. obtained from site-survey measurements). For each sample $i$, an ideal switching label $\delta_{\text{ref}}(i)\in\{0,1\}$ is generated according to the desired policy, where $\delta_{\text{ref}}(i)=1$ indicates BD-RIS OFF and $\delta_{\text{ref}}(i)=0$ indicates BD-RIS ON. We evaluate a discrete candidate set $\mathcal{P}$ of $P_{\text{fa}}$ values and select $P_{\text{fa}}^\star$ that maximizes $P_{\text{a}}$ on this dataset, as summarized in Algorithm \ref{algorithm1}.
\noindent
\begin{algorithm}[t]
\caption{Offline Calibration of $P_\text{fa}^\star$}\label{algorithm1}
\begin{algorithmic}[1]
\REQUIRE Offline calibration dataset $\mathcal{D}=\{1,\ldots,I\}$, candidate set $\mathcal{P}$ of $P_{\text{fa}}$, and data for testing
\ENSURE The optimal $P_{\text{a}}^\star$ and the corresponding $P_{\text{fa}}^\star$
\FOR{each offline sample $i \in \mathcal{D}$}
    \STATE Calculate offline reference switch state $\delta_{\text{ref}}(i)$;
\ENDFOR

\FOR{each $P_{\text{fa}} \in \mathcal{P}$}
    \STATE Extract data and calculate threshold by \eqref{eq:T_closedform1};
    \STATE Calculate offline real-time switch state $\delta_{\text{rw},P_{\text{fa}}}(i)$

    \FOR{each $i \in \mathcal{D}$}
        \STATE $d(i)_{P_\text{fa}} = 1\{\delta_\text{ref}(i) == \delta_{\text{rw},P_\text{fa}}(i)\}$
    \ENDFOR
    \STATE Count all successful matches $I_{\text{correct}} = \sum_{i=1}^{I} d(i)_{P_{\text{fa}}}$;
    \STATE Calculate total accuracy rate $P_{\text{a}} = \frac{I_{\text{correct}}}{I}$;
\ENDFOR

\STATE $P_{\text{fa}}^\star = \arg~\max_{P_{\text{fa}} \in \mathcal{P}} P_{\text{a}}$;
\STATE $P_{\text{a}}^\star = \max_{P_{\text{fa}} \in \mathcal{P}} P_{\text{a}}$;
\RETURN $P_{\text{a}}^\star$ and $P_{\text{fa}}^\star$;
\end{algorithmic}
\end{algorithm}
During the online operation, the selected $P_{\text{fa}}^\star$ is configured to the UE by feedback. The UE only needs to compute the decision threshold using \eqref{eq:T_closedform1} and \eqref{eq19} with the configured $P_{\text{fa}}^\star$, and perform a comparison between the measured pilot power $|y(l^\star)|^2$ and the threshold to decide the BD-RIS state.
 The proposed threshold design method significantly improves link state detection reliability in codebook-driven beamforming systems. Although the analytical threshold relies on the Rayleigh distribution, the sample quantile method is distribution-free but requires more pilots. Moreover, the selected beam, which follows a Rice distribution, has the same second-order statistical properties as the Rayleigh distribution, allowing the threshold to be readily obtained, as will be discussed in the Appendix.
\vspace{-0.1 in}

\section{Simulation Results and Analysis}
We first calibrate the parameters within a 3D region where the BS is located (in meters) at $(25, 30, 0)$ and the BD-RIS is deployed (in meters) at $(25, 25, 10)$. The UE positions are not fixed and selected regionally. The number of transmit antennas is set as $N_T=6$. The BD-RIS has $ M = 36 $ elements from both transmitting and reflecting sides and they are divided into $ G = 9 $ groups, each containing $ G_M = 4 $ elements from both sides. Gaussian white noise power is set to $-{70}$dB. Path Loss calculates as
$\alpha(G_1,G_2,d) = \sqrt{\frac{G_1 G_2 c^2}{(4\pi d f)^2}}$,
where $ c $ is the speed of light and the center frequency $ f = {2.9} $ \textrm{GHz}. The transmit and receive antennas are omnidirectional with gains $ G_T = 1$ and $ G_R = 1 $, respectively. RIS antennas are categorized into vertical $G_A = 2(r+1)\varpi^{r}$ and horizontal $G_A = 2(r+1)\phi^{r}$ polarizations.
The parameter $ r $ can be tuned for optimal gain; for general applicability, we set $ r = 1$. 
We set $d_1$, $d_2$, $d_3$ as the distance between the BS to the UE, the distance between the BS to BD-RIS, and the distance between BD-RIS to the UE, respectively.
Thus, we can obtain the path loss $\alpha_1=\alpha(G_T,G_R,d_1)$, $\alpha_2=\alpha(G_T,G_A,d_2)$, $\alpha_3=\alpha(G_A,G_R,d_3)$. 
The transmit power $ p $ is fixed to ${1}~$W.

\begin{figure}[t]
    \centering
    \includegraphics[width=0.45\textwidth]{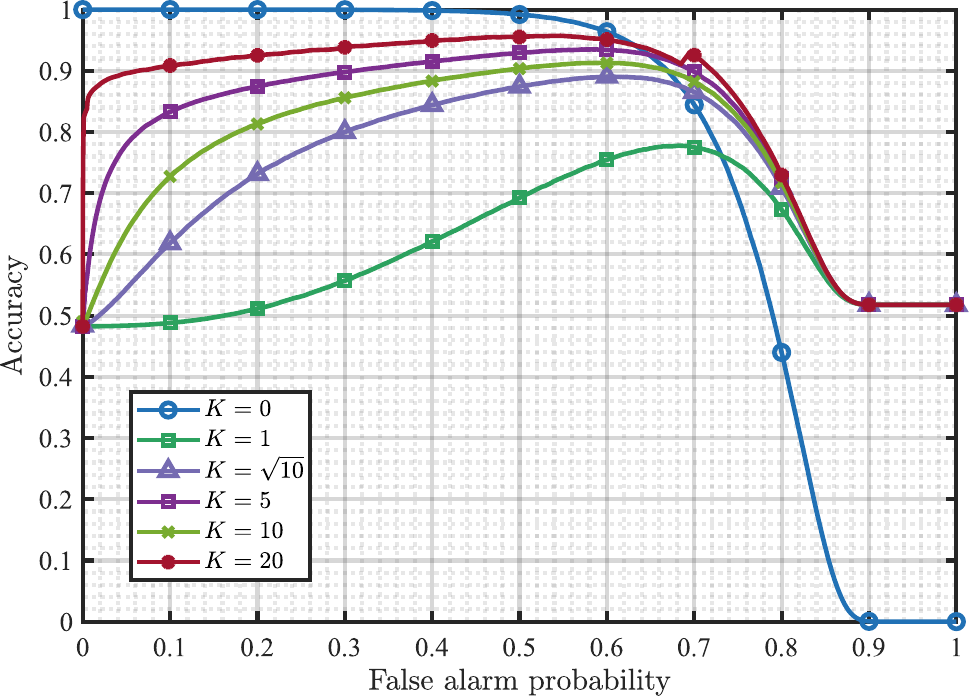}
    \caption{Detection accuracy for false alarm probability with different $K$.}\label{fvr}\vspace{-0.3 cm}
\end{figure}

To optimize the false alarm probability, the UE coordinates are distributed within a $20 \times 40 \times 10$ m$^3$ cuboid centered (in meters) at $(25, 25, 15)$. The cuboid is discretized via grid sampling to generate a reference table. To visualize the false alarm probability, we adopt a normalized format by introducing the number of antennas into the threshold, with $\tilde{P}_\text{fa}=(P_{\text{fa}})^{\frac{1}{N_T}}$. The detection accuracy corresponding to different false alarm probabilities is shown in Fig.~\ref{fvr}, from which we observe that the Rician factor $K$ impacts significantly the detection accuracy. Specifically, increasing the Rician factor $ K $ yields a higher probability of discrimination and thus a lower probability of false alarm.
Fortunately, regardless of the value of $K$, the highest detection accuracy occurs when the $\tilde{P}_\text{fa}$ is between 0.5 and 0.7. This allows us to set a common false alarm probability for various channel conditions to facilitate the proposed protocol. 

To evaluate the transmission performance, we use the effective throughput as the main metric. The proposed protocol is compared with two benchmark protocols.  \textit{1) Without RIS protocol:} In this case, the transmission relies only on the direct link and thus the transmission overhead is $L$. \textit{2) BD-RIS always-ON protocol:} In this case, the BS first attempts direct link communication using the codebook $\mathcal{W}$ to determine which mode the BD-RIS should work on (similar to Phase 1 in the proposed protocol). Then the downlink channel estimation is performed using the least squares method \cite{BD_RIS_CE} with an overhead $L + LC$. With channel estimate fed back from the user, the BS alternatively optimizes the BS beamformer and BD-RIS matrices $\mathbf{\Theta}_\mathsf{t}$ or $\mathbf{\Theta}_\mathsf{r}$. Therefore, the transmission overhead for this protocol is given by $2L+LC$. 

\begin{figure}[t]
    \centering
    \subfigure[$x$-axis]{\includegraphics[width = 0.48\linewidth]{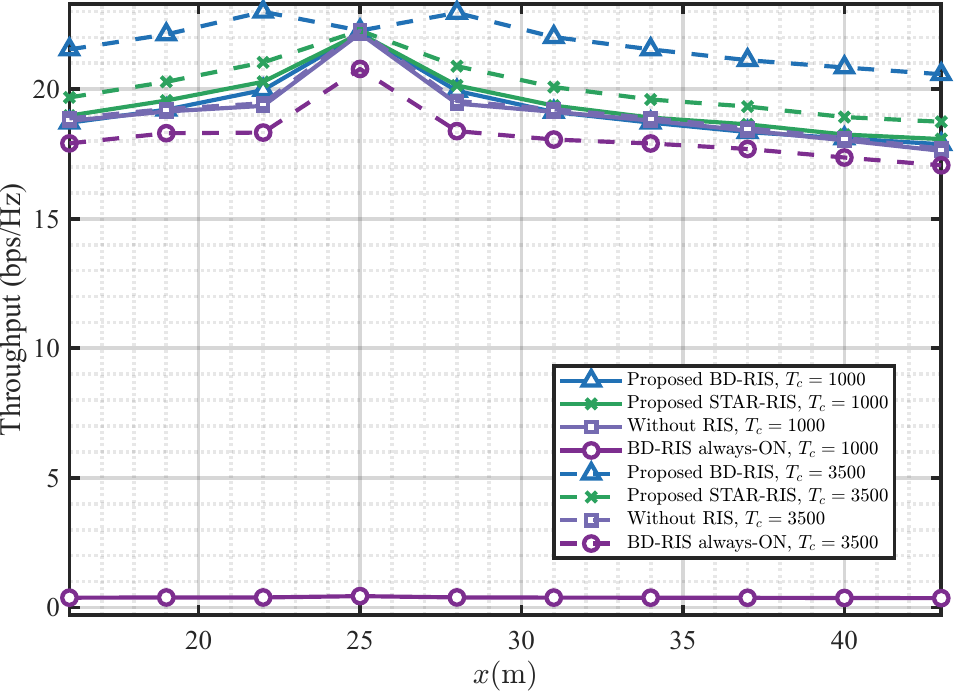}}
    \subfigure[$y$-axis]{\includegraphics[width = 0.48\linewidth]{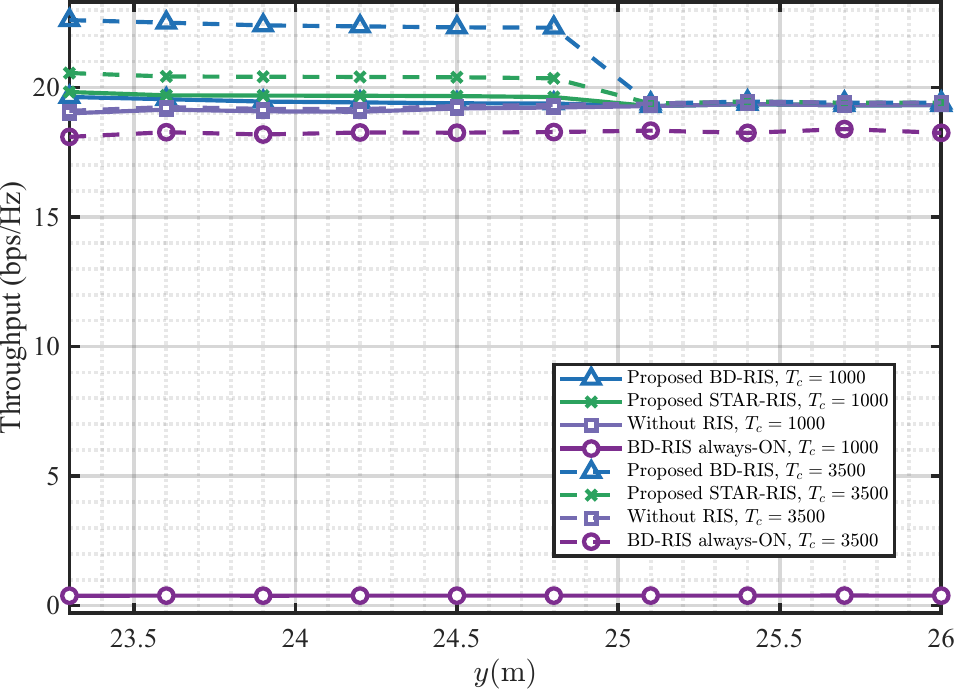}}
    \caption{Throughput versus different locations along (a) $x$-axis and (b) $y$-axis of the UE with and without the aid of BD-RIS ($G_M=4$) and STAR-RIS.}\label{fig:combined_power}\vspace{-0.3 cm}
\end{figure}

We assume that the UE moves along the $y$-axis from $(25,23.3,10)$ to $(25,26,10)$ and along the $x$-axis from $(16,25,10)$ to $(43,25,10)$. We also consider fast moving and slow moving scenarios that have respectively $T_c = 1000$ and $T_c = 3500$. The throughput is then computed for different state selection strategies and their corresponding switching decisions along the two paths as shown in Fig.~\ref{fig:combined_power}(a) and Fig.~\ref{fig:combined_power}(b).
Specifically, in Fig.~\ref{fig:combined_power}(a), scenarios with or without RIS obtain the same throughput in the point of $x = 25$ m due to the location setting.
From Fig. \ref{fig:combined_power}(b) we observe that, with a relatively large $T_c$, applying the proposed state switching strategy of BD-RIS can effectively improve the throughput compared to the benchmark schemes. The BD-RIS always-ON protocol is not efficient, especially when $T_c$ is small, since the large overhead of channel estimation overwhelms the beamforming gain. 

\section{Conclusion}

This paper studies a novel transmission protocol for a single-user MISO system aided by a BD-RIS, which aims to reduce overhead, simplify beamforming, and improve system performance. 
Specifically, a state-switching strategy of BD-RIS is proposed to avoid unnecessary activation of BD-RIS while maintaining robust transmission performance. This novel strategy is supported by a decision threshold design. 
It provides a practical solution for adaptive optimization and dynamic resource allocation in future wireless systems, especially suitable for dynamic environments with varying channel conditions.  

\begin{appendix}[Derivation of (\ref{eq:T_closedform1}) When LoS Exists]\label{app}

We start by copying the received signal during the training process as
\begin{equation}\label{LoS_receivetrain}
y(l) = \sqrt{\frac{Kp}{K+1}}\mathbf{h}_{\text{L}}^{\mathsf{H}}\bm{w}(l)s(l)  + \sqrt{\frac{p}{K+1}} \mathbf{h}_{\text{N}}^\mathsf{H}\bm{w}(l)s(l)  + n(l), \forall l,
\end{equation}
where we assume $s(1) = \ldots = s(L) = 1$. Since $\mathbf{h}_\mathrm{N}\sim\mathcal{CN}(\mathbf{0},\alpha_1^2\mathbf{I}_{N_T})$ and $n(l)\sim\mathcal{CN}(0,\sigma^2)$,
we can obtain the first-order moment as
\vspace{-0.05in}\begin{equation}\label{1stoy}
\mathbb{E}\{y(l)\} = \sqrt{\frac{Kp}{K+1}} \mathbf{h}_{\text{L}}^\mathsf{H}\bm{w}(l), \forall l,
\end{equation}
and the second-order expectation as
\vspace{-0.05in}\begin{equation}\label{2edoy}
\mathbb{E}\{|y(l)|^2\} = \frac{K}{K+1} p \mathbf{h}_\mathrm{L}^\mathsf{H}\bm{w}(l)\bm{w}(l)^\mathsf{H}\mathbf{h}_\mathrm{L} + \frac{1}{K+1} p \alpha_1^2  + \sigma^2, \forall l.
\end{equation}
The variance of the received signal is thus
\begin{align}\label{eq19}
\text{var}(y(l)) =&\mathbb{E}\{|y(l)|^2\} - \mathbb{E}\{y(l)^*\}\mathbb{E}\{y(l)\}\nonumber\\
=&\frac{p}{K+1} \alpha_1^2  + \sigma^2 = \epsilon,
\end{align}
which is exactly the information required to calculate the threshold in (\ref{eq:T_closedform1}). Note that the variance of the received signals should be estimated in practice to facilitate the whole procedure of the proposed protocol. The fact that received signals share the same variance regardless of the existence of LoS components can potentially simplify the required estimation, thereby supporting the feasibility of the proposed threshold detection and protocol.
\end{appendix}    

\section*{Acknowledgment}

This work is funded by the National Natural Science Foundation of China (grant no. 62501509) and the Natural Science Foundation of Guangdong Province (grant no. 2026A1515011048).

\bibliographystyle{IEEEtran}
\bibliography{refs}

@ARTICLE{RIS2.0,
  author={Li, Hongyu and Nerini, Matteo and Shen, Shanpu and Clerckx, Bruno},
  journal={IEEE Commun. Surveys \& Tuts.}, 
  title={A Tutorial on Beyond-Diagonal Reconfigurable Intelligent Surfaces: Modeling, Architectures, System Design and Optimization, and Applications}, 
  year={2026},
  volume={28},
  number={},
  pages={4086-4126},
  doi={10.1109/COMST.2025.3647003}}

@article{STAR,
  title={Simultaneously transmitting and reflecting {(STAR) RIS} aided wireless communications},
  author={Mu, Xidong and Liu, Yuanwei and Guo, Li and Lin, Jiaru and Schober, Robert},
  journal={IEEE Trans. Wireless Commun.},
  volume={21},
  number={5},
  pages={3083--3098},
  year={2021},
  publisher={IEEE}
}

@ARTICLE{BD_RIS_CE,
  author={Li, Hongyu and Shen, Shanpu and Zhang, Yumeng and Clerckx, Bruno},
  journal={IEEE Trans. Signal Process.}, 
  title={Channel Estimation and Beamforming for Beyond Diagonal Reconfigurable Intelligent Surfaces}, 
  year={2024},
  volume={72},
  number={},
  pages={3318-3332},
  doi={10.1109/TSP.2024.3424229}}

@ARTICLE{CE_survey,
  author={Zheng, Beixiong and You, Changsheng and Mei, Weidong and Zhang, Rui},
  journal={IEEE Commun. Surveys \& Tuts}, 
  title={A Survey on Channel Estimation and Practical Passive Beamforming Design for Intelligent Reflecting Surface Aided Wireless Communications}, 
  year={2022},
  volume={24},
  number={2},
  pages={1035-1071},
  doi={10.1109/COMST.2022.3155305}}

@ARTICLE{CE_RIS,
  author={Chen, Jie and Liang, Ying-Chang and Cheng, Hei Victor and Yu, Wei},
  journal={IEEE Trans. Wireless Commun.}, 
  title={Channel Estimation for Reconfigurable Intelligent Surface Aided Multi-User mmWave {MIMO} Systems}, 
  year={2023},
  volume={22},
  number={10},
  pages={6853-6869},
  doi={10.1109/TWC.2023.3246264}}

@ARTICLE{CE1_RIS,
  author={Wang, Zhaorui and Liu, Liang and Cui, Shuguang},
  journal={IEEE Trans. Wireless Commun.}, 
  title={Channel Estimation for Intelligent Reflecting Surface Assisted Multiuser Communications: Framework, Algorithms, and Analysis}, 
  year={2020},
  volume={19},
  number={10},
  pages={6607-6620},
  doi={10.1109/TWC.2020.3004330}}

@ARTICLE{FB,
  author={You, Changsheng and Zheng, Beixiong and Zhang, Rui},
  journal={IEEE Wireless Commun. Lett.}, 
  title={Fast Beam Training for {IRS}-Assisted Multiuser Communications}, 
  year={2020},
  volume={9},
  number={11},
  pages={1845-1849},
  doi={10.1109/LWC.2020.3005980}}

@ARTICLE{BT,
  author={Wang, Peilan and Fang, Jun and Zhang, Weizheng and Chen, Zhi and Li, Hongbin and Zhang, Wei},
  journal={IEEE Wireless Commun.}, 
  title={Beam Training and Alignment for {RIS}-Assisted Millimeter-Wave Systems: State of the Art and Beyond}, 
  year={2022},
  volume={29},
  number={6},
  pages={64-71},
  doi={10.1109/MWC.006.2100517}}

@ARTICLE{HR,
  author={Wang, Jinghe and Tang, Wankai and Jin, Shi and Wen, Chao-Kai and Li, Xiao and Hou, Xiaolin},
  journal={IEEE Trans. Commun.}, 
  title={Hierarchical Codebook-Based Beam Training for {RIS}-Assisted mmWave Communication Systems}, 
  year={2023},
  volume={71},
  number={6},
  pages={3650-3662},
  doi={10.1109/TCOMM.2023.3251374}}

@ARTICLE{match_filter,
  author={Lo, T.K.Y.},
  journal={IEEE Trans. Commun.}, 
  title={Maximum ratio transmission}, 
  year={1999},
  volume={47},
  number={10},
  pages={1458-1461},
  doi={10.1109/26.795811}}

@article{lehmann1993fisher,
  title={The Fisher, Neyman-Pearson theories of testing hypotheses: one theory or two?},
  author={Lehmann, Erich L},
  journal={J. American statistical Association},
  volume={88},
  number={424},
  pages={1242--1249},
  year={1993},
  publisher={Taylor \& Francis}
}

\end{document}